\documentclass[aps,prd,showpacs,superscriptaddress,letterpaper,preprintnumbers,asmmath,amssymb]{revtex4}
\usepackage{amssymb,amsmath,amsbsy}
\usepackage{mathrsfs}
\usepackage[dvips]{graphics}
\usepackage{epsfig}

\def\be{\begin{eqnarray}}
\def\ee{\end{eqnarray}}
\def\bea{\begin{eqnarray*}}
\def\eea{\end{eqnarray*}}

\def\centeron#1#2{{\setbox0=\hbox{#1}\setbox1=\hbox{#2}\ifdim
\wd1>\wd0\kern.5\wd1\kern-.5\wd0\fi
\copy0\kern-.5\wd0\kern-.5\wd1\copy1\ifdim\wd0>\wd1
\kern.5\wd0\kern-.5\wd1\fi}}
\def\ltap{\;\centeron{\raise.35ex\hbox{$<$}}{\lower.65ex\hbox{$\sim$}}\;}
\def\gtap{\;\centeron{\raise.35ex\hbox{$>$}}{\lower.65ex\hbox{$\sim$}}\;}

\newcommand{\newc}{\newcommand}
\newc{\qbar}{{\overline q}}
\newc{\Kahler}{K\"ahler }
\newc{\deltaGS}{\delta_{\rm GS}}

\begin{document}
\preprint{
\vbox{\vspace*{2cm}
      \hbox{UCI-TR-2011-13}
      \hbox{July, 2011}
}}
\vspace*{3cm}

\title{Higgs Decays to Unstable Neutrinos:\\ Collider Constraints from Inclusive Like-Sign Dilepton Searches}
\author{Linda M. Carpenter}
\author{Daniel Whiteson}

\affiliation{Department of Physics and Astronomy  \\
   University of California, Irvine, CA 92697 \\
\vspace{1cm}}

\begin{abstract}

We study the pair production of fourth generation neutrinos from the
decay of an on-shell Higgs produced by gluon fusion.   In a fourth
generation scenario, the Higgs branching fraction into fourth
generation neutrinos may be quite large.  In the case that the
unstable heavy neutrinos are a mixed Majorana and Dirac state,
neutrinos pair-produced from Higgs decay will yield a substantial
number of like-sign dilepton events. In this article we use inclusive
like-sign dilepton searches from hadron colliders to constrain the
theoretical parameter space of fourth generation leptons. 

\end{abstract}
\pacs{}

\maketitle

\section{Introduction}

A fourth generation is among the simplest possibilities for new
physics at the weak scale. If a fourth generation exists,  recent work
has shown that bounds on unstable fourth generation leptons may be
well under 100 GeV, making the leptons the lightest new states.  Heavy
unstable  fourth generation neutrinos would decay through the process $N_1\rightarrow \ell W$.  Fourth generation neutrinos may be light, LEP placed mass bounds under 100 GeV  \cite{Achard:2001qw} \cite{Abulencia:2007ut}.  Most generally, fourth generation lepton sectors may have neutrinos with mixed Dirac and Majorana masses and, in this case, mass bounds on the lightest neutrinos may be 62.1, 79.9 or 81.8 GeV, depending on whether the final state lepton is $\tau$, $\mu$, or e \cite{Carpenter:2010dt}.

If a fourth generation exists, Higgs physics may provide a powerful
handle for constraining these models.  The dominant Higgs production
mode at both Tevatron and LHC is gluon fusion, $gg\rightarrow h$.
This process proceeds through a heavy-quark loop and is substantially
enhanced by the presence of fourth generation quarks.  The enhancement
is largely independent of quark mass and gives an increase in Higgs production
of approximately a factor of eight compared to production in the Standard Model, see for example \cite{Kribs:2007nz} \cite{Li:2010fu}.

The presence of fourth generation particles also greatly effects the
Higgs branching fractions.  For SM Higgs masses up to 160 GeV, the
Higgs decay width is dominated by decay to bottom quarks; however,
this proceeds through a Yukawa coupling that is not extremely large.
If other heavier states exist in the theory, they may easily dominate
and require us to look for the Higgs in non-standard channels.  For
example, many recent 'hidden Higgs' scenarios yield highly altered
Higgs branching fractions and a variety of standard Higgs final
states, for example see \cite{hidden}-\cite{hidden5}.  A fourth-generation lepton
sector also offers new channels for Higgs decays.  Fourth generation
neutrinos with both Dirac and Majorana masses will have significant
couplings to the Higgs depending on the Dirac mass component.  In fact, as long as the Dirac mass parameter is larger than the bottom mass, the Higgs decay width to neutrinos will dominate that of bottoms for any sufficiently heavy Higgs mass.  In the window of Higgs masses between roughly 120 and 160 GeV, the dominant branching fraction of the Higgs may be to heavy neutrinos.  For larger values of the Higgs mass, where the decay channel into on-shell electro-weak gauge bosons is open, the branching fraction into neutrinos is still significant, remaining above 10 percent for Higgs masses up to 200 GeV and above one percent up to 500 GeV.

In principle fourth generation neutrinos may be stable or unstable,
leading to very different Higgs decay signatures. Stable Majorana
neutrinos may be quite light, under 50 GeV \cite{pdg} \cite{Abreu:1991pr} \cite{Lenz:2010ha} \cite{Carpenter:2010sm}, and the
possibility of light stable neutrinos as a Higgs decay channel has
been recently explored \cite{Belotsky:2002ym} \cite{Keung:2011zc}.  Unstable Majorana
neutrinos may decay into a standard model charged lepton and a $W$
boson, $N_1\rightarrow W\ell$.  If this decay is flavor-democratic, or
is dominated by $\tau$ decays, the heavy neutrino may be as light as
61.2 GeV.  Since a Majorana neutrino may decay into a SM lepton of
either sign, any process which pair-produces these fourth generation
neutrinos produces many like-sign dileptons \cite{Rajaraman:2010ua} \cite{Rajaraman:2010wk}
.  As like-sign dileptons
are a clear, low-background signature at hadron colliders, the
possibility of Higgs decays in this channel is quite interesting.  Previous works have proposed searches for heavy neutrino pairs produced from a Z and which decay into like-sign dileptons.  In
this case, the signal of like-sign dileptons plus jets become an interesting new channel in which to look for the Higgs.  If the
Higgs production as well as the branching fraction to fourth
generation neutrinos are large enough,  a simple inclusive like-sign
dileptons search may be used to constrain large parts of fourth
generation parameter space - this possibility is the focus of this
paper.  We will use published inclusive same-sign dilepton analyses
from LHC and the Tevatron to place constraints on various fourth generation lepton scenarios.

This paper is organized as follows: Section 2 reviews fourth
generation neutrino masses and couplings, Section 3 discusses Higgs
production and branching fractions in fourth generations scenarios,
Section 4 analyses inclusive like-sign dilepton searches at hadron colliders, and Section 5 concludes.

\section{Review of Fourth Generation Neutrino Masses}

In the most general case, fourth generation neutrinos may have both a Dirac and a Majorana mass.  Note that here we use the notation of \cite{Katsuki:1994as},  where the Lagrangian is

\bea
L =  m_D\overline{L_4}N_R +M_N N^2+\frac{m_D}{v}H\overline{L_i}N_R
\eea

The mass matrix is given by,
\begin{eqnarray}
{\cal L}_m=-{1\over 2}\overline{(Q_R^c
N_R^c)}\left(\begin{array}{cc}0&m_D\\m_d&M\end{array}\right)
\left(\begin{array}{c}Q_R\\ N_R\end{array}\right)+h.c.
\end{eqnarray}
where $\psi^c=-i\gamma^2\psi^*$.
There are then two Majorana neutrinos with different mass eigenvalues:
\bea
 \nonumber M_1=-(M/2)+ \sqrt{m_D^2+{M^2/4}} \\ M_2=(M/2)+ \sqrt{m_D^2+{M^2/4}}
\eea

The mass eigenstates can be expressed in terms of the left-right eigenstates

 \bea
 N_1=\cos\theta Q_L^c+\sin\theta N_R+\cos\theta  Q_L+\sin\theta N_R^c
 \\
 N_2=-i\sin\theta Q_L^c+i\cos\theta  N_R+i\sin\theta Q_L-i\cos\theta  N_R^c
  \eea

\noindent
  where we have defined the mixing angle
\bea
\nonumber \tan\theta=m_1/m_D
\eea

The Higgs couples to the neutrino mass eigenstates with coupling proportional to powers of the neutrino mixing angle.
The Higgs coupling to the lightest neutrino pair is given by
\bea
\frac{m_D}{v_h}H N_1 N_1 sin(2\theta).
\eea
The neutrino mixing angle varies between $\pi / 4$ for pure Dirac-type
neutrinos and $\pi/2$ for Majorana-type. We see that in the limit
where the neutrinos are pure Majorana, $m_D$ approaches zero, the mixing
angle $\theta$ approaches $\pi/2$, and the Higgs decouples from the neutrinos as required, while for pure Dirac states the coupling is maximal.

\section{Higgs Production and Branching Fractions}
\begin{figure}[h]
\centerline{\includegraphics[width=2.5 in]{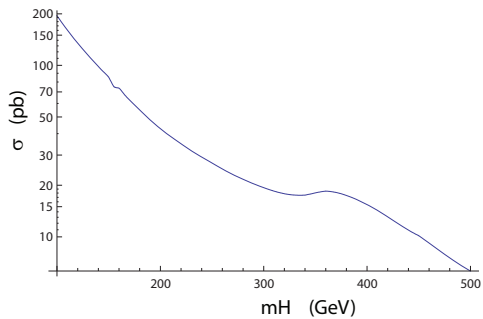}}
\caption{Dependence of $gg\rightarrow h$ production on Higgs mass, including
  effects of fourth generation quarks, at LHC with $\sqrt{s}=7$ TeV. }
\label{fig:2pgm}
\end{figure}

\begin{figure}[h]
\centerline{\includegraphics[width=8 cm]{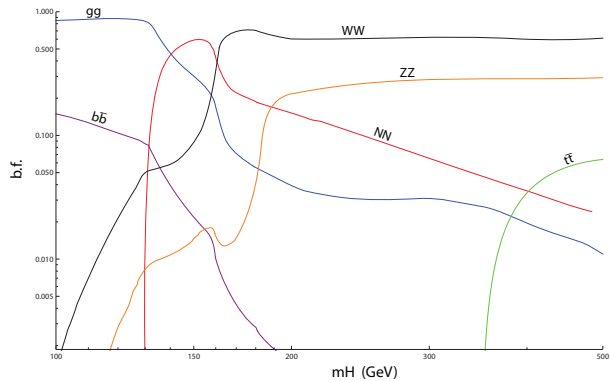}}
\caption{Dependence of Higgs branching fractions on mass for the benchmark point $m_{n_1} = 63 GeV$, $m_{n_2}=300 GeV$. }
\label{fig:2pgm}
\end{figure}

At hadron colliders, the largest Higgs production rate comes from gluon fusion.  The Higgs-gluon coupling is generated by loops of the heavy quarks.  The $gg\rightarrow h$ production cross section is given by

\bea
\label{production}
\sigma (p\bar{p} \to h) &=& \frac{1}{16}\Gamma (h\to gg) \frac{16 \pi^2}{s m_h} \>\int_{m_h^2/s}^1 \frac{dx}{x} g(x) g(\frac{m_h^2}{s  x}),
\eea

where $\Gamma (h\to gg)$ is the Higgs to gluon decay width

\bea
 \Gamma(h\rightarrow gg)=\frac{\alpha_s G_F m_h^3}{16 \sqrt{2} \pi^3}\Sigma_i( {\tau_i}(1+(1-\tau_i)f(\tau_i)))
\eea
with
\bea
\tau_i=\frac{4 m_{f_i}^2}{m_h^2},  f(\tau_i)= ({sin}^{-1}\sqrt{1/\tau_i})^2
\eea

where the index $i$ runs over heavy quark flavors.  Due to their large
Yukawa couplings, the additional  of fourth generation quarks enhances
the $h\rightarrow gg$ decay width substantially; the resulting decay
width is largely independent of the new heavy quark mass, and larger
than the Standard Model prediction by about a factor of 8. The
$gg\rightarrow h$ production cross-section in a fourth generation scenario for LHC is shown in Figure 1; note that the Higgs production remains above a picobarn for Higgs masses up to 500 GeV.

For Higgs masses below twice the $W$ or $Z$ masses, the most important
standard model Higgs branching fractions are b quarks and gluons. This
changes significantly with the addition of a fourth generation. The Higgs decay width into a Dirac particle  is given by

\bea
\Gamma_D = \frac{m_h y_D^2}{8 \pi}(1-\frac{4m_d^2}{m_h^2})^{3/2}
\eea

The branching ratio is proportional to the square of the Yukawa
coupling.  We see that the bottom, which normally dominates the Higgs
decay for low masses, does not have a particularly large Yukawa
coupling.  A fourth generation neutrino with substantial Dirac
component may easily overtake the bottom decay mode, see Figure 2.

The ratio of  $h \rightarrow N_1 N_1$ to $h \rightarrow b\overline{b}$
is
\bea
\frac{\Gamma(h\rightarrow n_1 n_1)}{\Gamma(h\rightarrow b\overline{b})}= \frac{m_{n1}^2 \sin(2\theta)}{m_b^2} \frac{(1-\frac{4m_{n1}^2}{mh^2})^{3/2}}{(1-\frac{4m_b^2}{mh^2})^{3/2}}
\eea

Notice, however, that the coupling is proportional to the mixing
angle.  Only when the neutrino is in the deep Majorana limit where
$\theta\rightarrow \pi/2$ does the Higgs branching fraction to
neutrinos drop substantially. Figure 3 shows the effect of mixing
angle on Higgs branching fraction. These are contour plots over the
$N_1,N_2$ mass plane for two benchmark Higgs mass values.  The
branching fraction drops as one increases $N_2$ mass and enters the
Majorana region.

\begin{figure}[h]
\begin{center}$
\begin{array}{cc}
\includegraphics[width=2.0in]{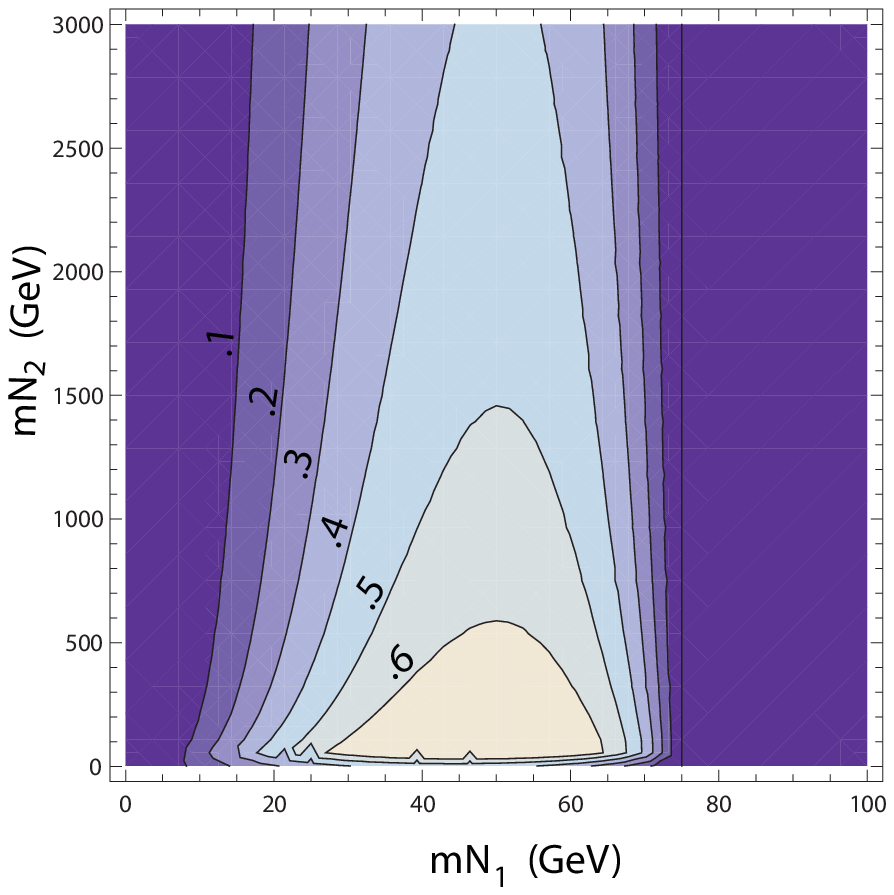} &
\includegraphics[width=2.0in]{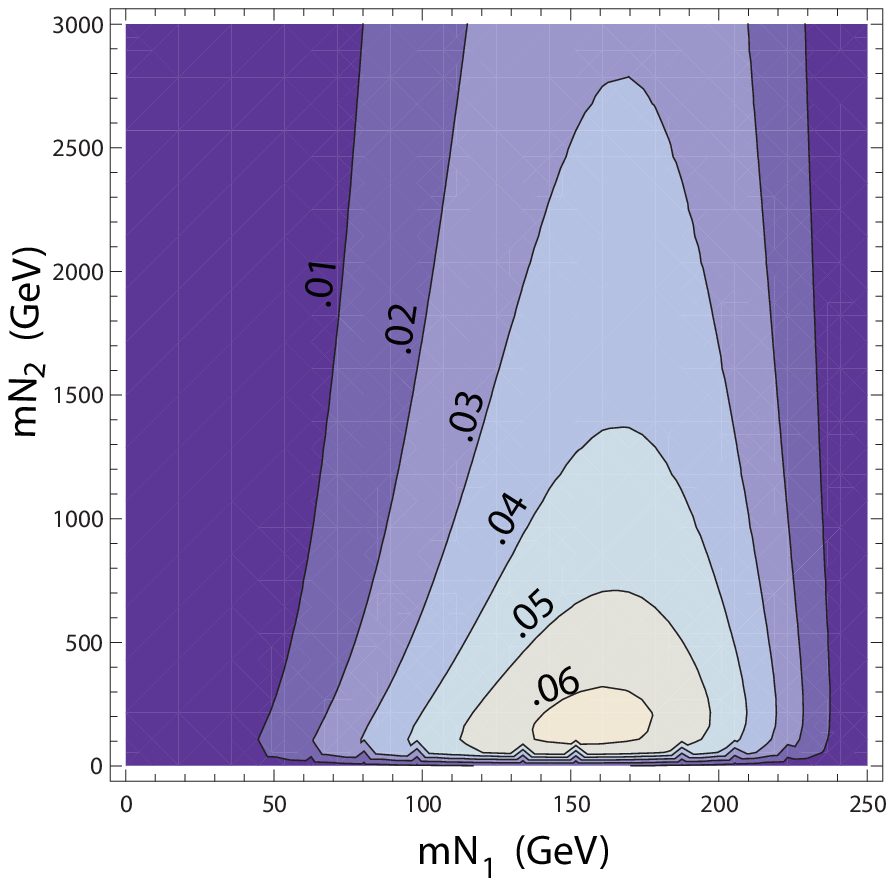}
\end{array}$
\end{center}
\caption{ Higgs branching fraction $h\rightarrow N_1 N_1$ Higgs mass
  values of 150 GeV (left) and 400 GeV (right). }
\end{figure}

In the Standard Model, the next largest branching fraction comes from
Higgs decaying to gluons.  The decay width of Higgs to gluons is given
above. With the addition of a fourth generation, the branching
fraction of Higgs to gluons is substantially larger than it is in the
Standard Model and overtakes that of $h \rightarrow b\overline{b}$ for
low masses. However, in the region just above twice the mass of the
lightest fourth generatio neutrino, $h\rightarrow N_1 N_1$ is the {\it
  dominant} decay mode, see Figure 2. Once the Higgs becomes large enough to  decay into a pair of on-shell
electroweak gauge bosons, this mode dominates the Higgs branching
fraction.  However, the decay width into fourth generation neutrinos
remains larger than all other channels.    The Higgs
branching fraction to neutrinos remains appreciable, above 10 percent,
even for Higgs masses out to 200 GeV and above a percent up to Higgs
masses of 500 GeV.


In the heavy Higgs region the most sensitive search channel is the
$h\rightarrow ZZ$ mode. However, to suppress the multi-jet background
at least one leptonic decaying $Z$ is required, which significantly
reduces the cross-section due to the small $Z\rightarrow \ell\ell$
branching ratio.  If the Higgs decay to fourth-generation neutrinos is
very distinctive, it may remain an important channel for high mass
Higgs searches.


\section{Collider Bounds}
\subsection{LHC}
We consider Higgs production from gluon fusion and decay to a pair of
light fourth generation neutrinos, where the lightest neutrino is unstable
and decays via the process $N_1 \rightarrow W\ell$, giving

\[ gg\rightarrow h\rightarrow N_1N_1\rightarrow WW\ell\ell\rightarrow
\ell\ell + qq'+qq' \]

  The $N_1 \rightarrow W\ell$ decay rate is determined by the values
  of a four-generation lepton mixing matrix.  In the simple case that we consider here, we allow only a
single decay rate to be non-zero, so that the fourth generation
neutrino decays entirely to a single flavor of lepton.

 Because the $N_1$ state is Majorana, it may decay to a final state
 lepton of either sign with equal probability: thus the decay results
 in like-sign dileptons half of the time.  Therefore, there is a
 significant rate of Higgs production with decay into states with high
 $p_T$ like-sign dileptons, a low-background signature.

\begin{figure}[h]
\centerline{\includegraphics[width= 3.0 in]{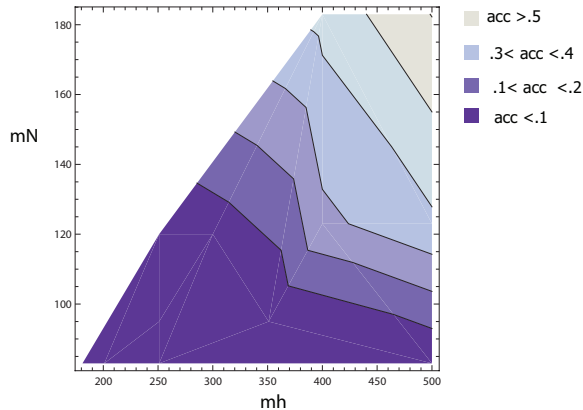}}
\caption{Plot of ATLAS fiducial region acceptance (defined in text) in the $m_h$, $N_1$ mass-plane. }
\label{fig:2pgm}
\end{figure}

ATLAS has completed an inclusive like-sign dilepton search which looks
for like sign muons of high invariant mass in 34 pb$^{-1}$ of data \cite{Aad:2011vj}.
Even with a small data set, ATLAS has set limits on the cross-section
of like-sign dimuon events at  $< 170$ pb in a simple fiducial region.
We calculate the predicted cross-section in the ATLAS fiducial region
for fourth-generation Higgs decays to to constrain fourth generation
parameter space with the inclusive like sign dilepton data.  Higgs
events were generated using MADGRAPH \cite{Alwall:2007st} decayed with
BRIDGE \cite{Meade:2007js} and showered with PYTHIA \cite{pythia}.  The
fiducial region requires two like-sign muons with

\begin{itemize}
\item{$p_T >$20 GeV and $\eta_{\mu} < 2.5$ }
\item{isolation cone of $R>0.4$ between muons and quarks or gluons}
\item{di-muon invariant mass of  $>110$ GeV }
\end{itemize}

The acceptance for  $h \rightarrow n_1 n_1 \rightarrow \ell\ell WW$ in
this fiducial region is shown in Figure 4.  Notice that the acceptance becomes large when the Higgs mass is large, due to the high invariant mass cut on the like-sign leptons.
 The acceptance also falls as the $N_1$ mass becomes lighter and the
 final state lepton becomes soft.

\begin{figure}[h]
\begin{center}$
\begin{array}{cc}
\includegraphics[width=3.0in]{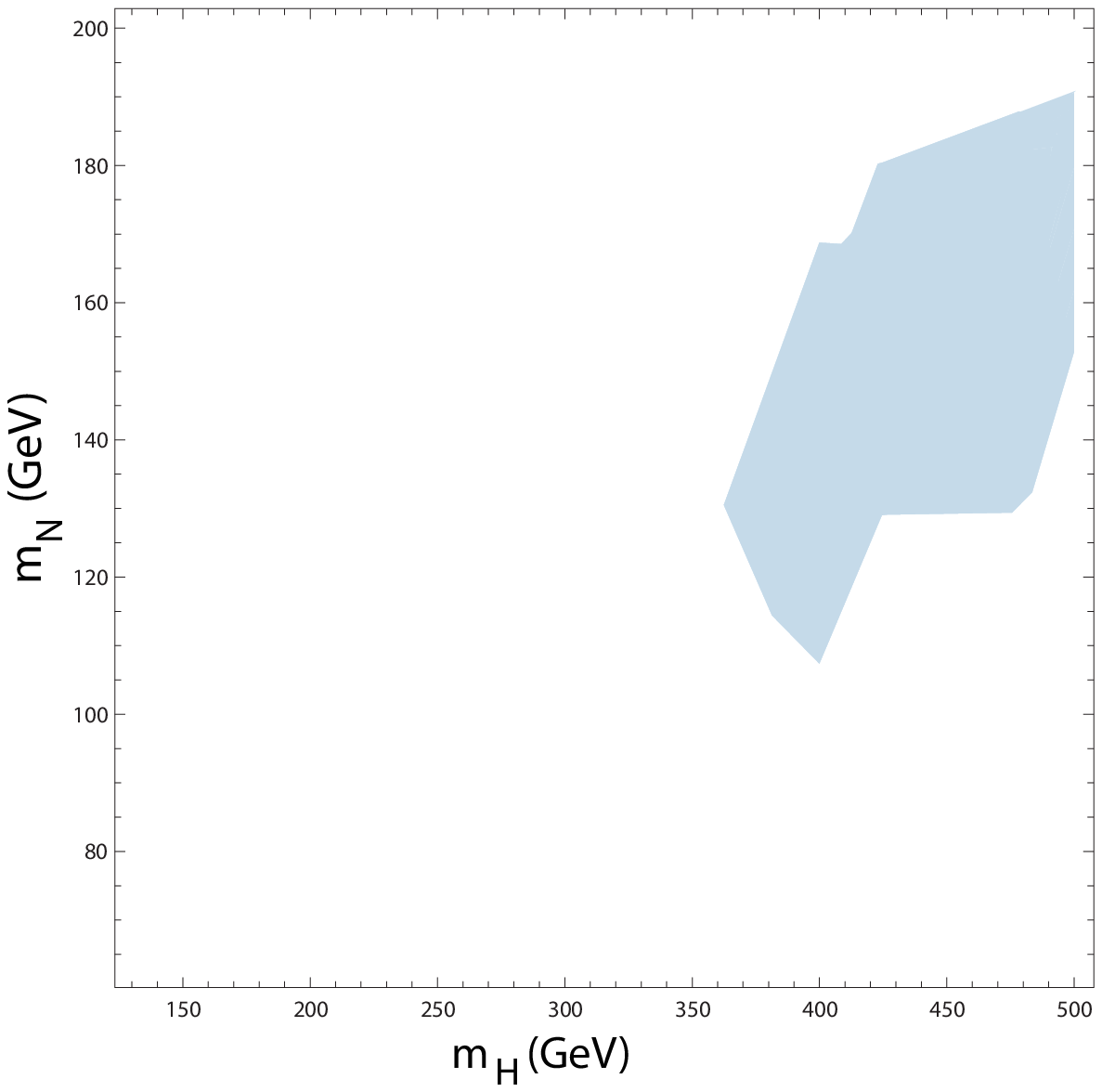} &
\includegraphics[width=3.05in]{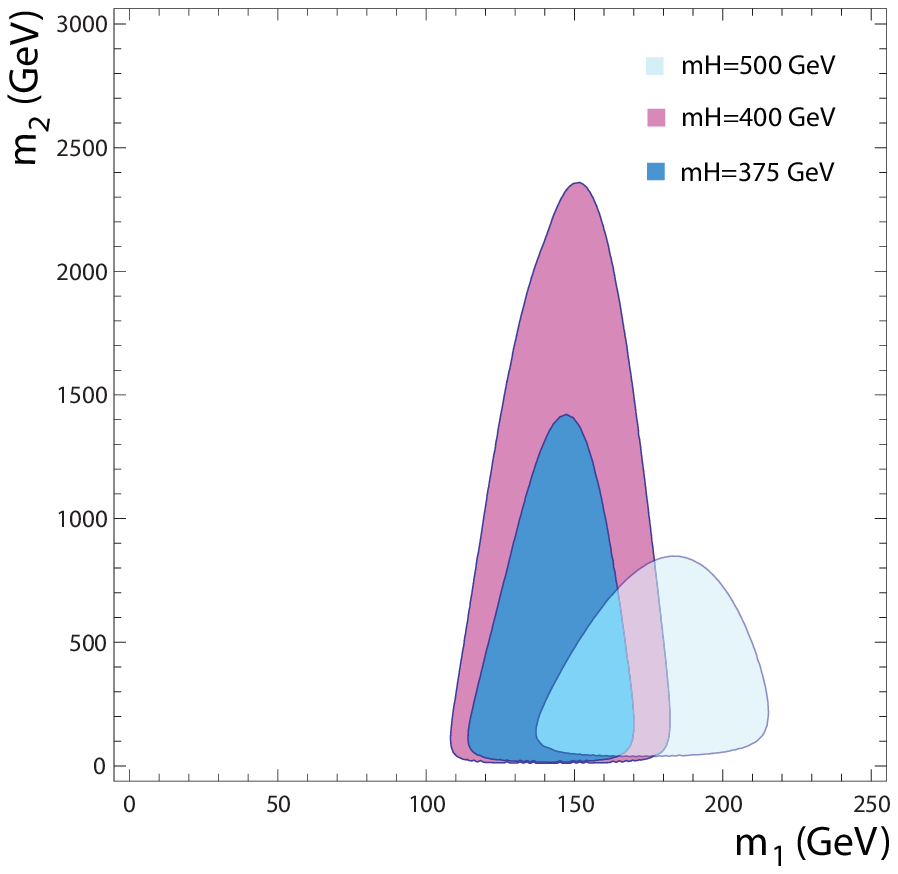}
\end{array}$
\end{center}
\caption{Exclusions at 95 percent confidence level from the ATLAS
  inclusive like-sign dilepton search in (left)the $N_1$, $m_h$ mass
  plane for the value $m_{N_2}=300$ GeV, and (right) the $N_1$,$N_2$ mass plane for Higgs masses of 375 ,400, and 500 GeV. }
\end{figure}

Figure 5 shows the expected exclusion at 95 percent confidence  level
in the Higgs, $N_1$ mass plane from ATLAS data for the fixed value
$m_{N_2}= 300$ GeV.  Figure 5 gives an exclusion in the $N_1, N_2$ mass plane shown for different values of the Higgs mass.  For large mixing between neutrinos, light neutrino masses between 100 and 200 GeV can be excluded.  Notice that regions of parameter space are ruled out as long as the Higgs branching fraction to neutrinos remains large: Again, when $N_2$ increases, the neutrino becomes Majorana-like and decouples from the Higgs, thus exclusions can no longer be made.

\subsection{Tevatron}
The CDF experiment has recently completed an inclusive like-sign
dilepton analysis using 6.1 $fb^{-1}$ of data \cite{CDFnote}.  Though
Tevatron's $gg\rightarrow h$ production cross section is smaller than
LHC's, it is still possible to use the inclusive like-sign dilepton
search to constrain fourth generation parameter space for Higgs masses
under 200 GeV. First of all, in a fourth generation scenario, Higgs
production from gluon fusion is substantially enhanced.  In addition,
in the regime of lighter Higgs masses, the lightest Majorana neutrinos
occupy a significant or even dominant portion of the Higgs branching
fraction, and we thus expect many like-sign dilepton events from Higgs
production.  In addition, Tevatron's inclusive like-sign dilepton
search did not require the large invariant mass cut, increasing the
sensitivity to the fourth generation Higgs decays where the neutrino
is light.

We consider again the process $gg\rightarrow h\rightarrow N_1N_1
\rightarrow WW\ell\ell\rightarrow \ell\ell + qq'+qq'$.  CDF's search
considered both electrons and muons in the final state.  Note that the
weakest bound in fourth generation lepton scenario occurs when the
fourth generation neutrino decay, $N_1\rightarrow W\ell$, is flavor
democratic: then the lightest allowed neutrino mass is 61.2 GeV.  We
expect that we will be able to improve the bound using the inclusive
like-sign dilepton search only in the flavor democratic neutrino decay
scenario.  To calculate the acceptance for CDF's cuts, we generated
events in {\sc madgraph} and showered them with {\sc pythia}.  We use CDF's inclusive like sign dilepton cuts:

\begin{itemize}
\item{2 same-sign leptons }
\item{the most energetic lepton $p_T >$20 GeV   }

\item{next most energetic lepton $p_T >$10 GeV   }
\item{$|\eta_{\ell}| < 1.1$ }
\item{veto of like-sign electrons with invariant masses 86-96 GeV}
\item{in tri-lepton events, veto of opposite-sign leptons with invariant masses 86-96 GeV }
\end{itemize}

 \begin{figure}[h]
\centerline{\includegraphics[width=2.5 in]{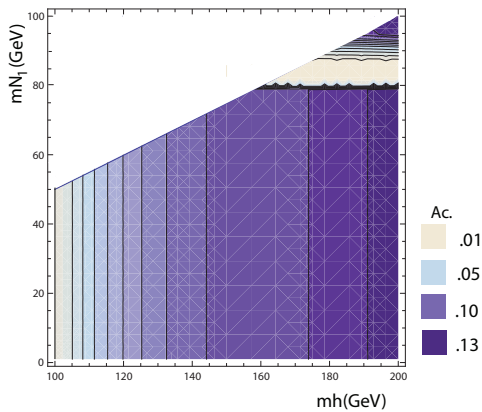}}
\caption{Acceptance of CDF's like-signed dilepton selection in the $m_h$, $N_1$ mass-plane. }
\label{fig:2pgm}
\end{figure}

Applying the CDF selection to our simulated events, we find
efficiences ranging up to 13\%, see Figure 6. We see that the
efficiency increases as the Higgs mass increases, as the decay
products of the Higgs become harder.  In addition, we see a decrease
in the efficiency for values of the $N_1$ mass above the $W$ mass.  This is expected as the lepton in the heavy neutrino decay  $n_1\rightarrow W\ell$ will be extremely soft for neutrino masses just above the $W$ mass.

 Backgrounds for this search are quite small and come from diboson
 production;$WZ$ and $ZZ$, as well as from misidentified leptons due
 to jets in top quark pair production or $W$+jet events.  Higgs
 branching fractions into neutrinos in this range may be quite large,
 as one can see from Figures 2 and 3.  We may expect to be able to
 rule out the existence of very light neutrinos at 95 percent
 confidence level in the case that decays to leptons are flavor
 democratic.  Figure 7 shows exclusions at 95 confidence level in the neutrino mass plane for 4 benchmark values of the Higgs mass.  Note that for Higgs masses up to $2m_W$, we exclude $N_1$ masses below the $W$ mass, unless we are into the very deep Majorana region.  Once the Higgs mass gets larger, the branching fraction to neutrinos is much less, and we can no longer rule out substantial pieces of parameter space.

 \begin{figure}[h]
\begin{center}$
\begin{array}{cc}
\includegraphics[width=1.5in]{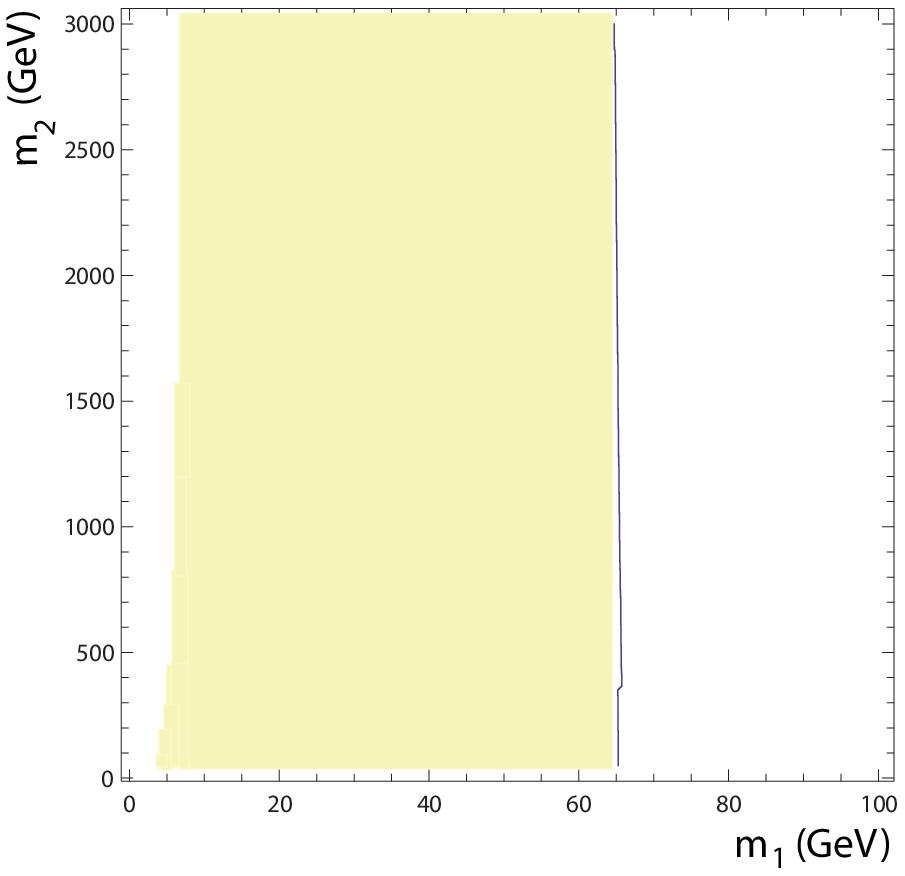} &
\includegraphics[width=1.5in]{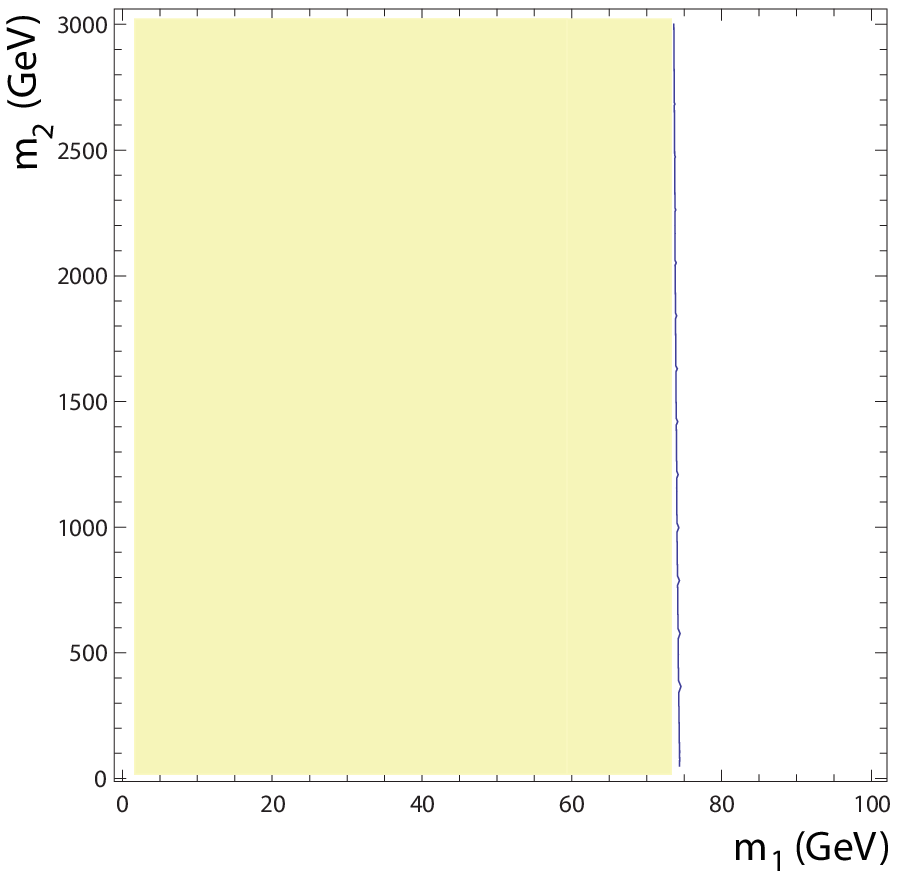} \\ \includegraphics[width=1.5in]{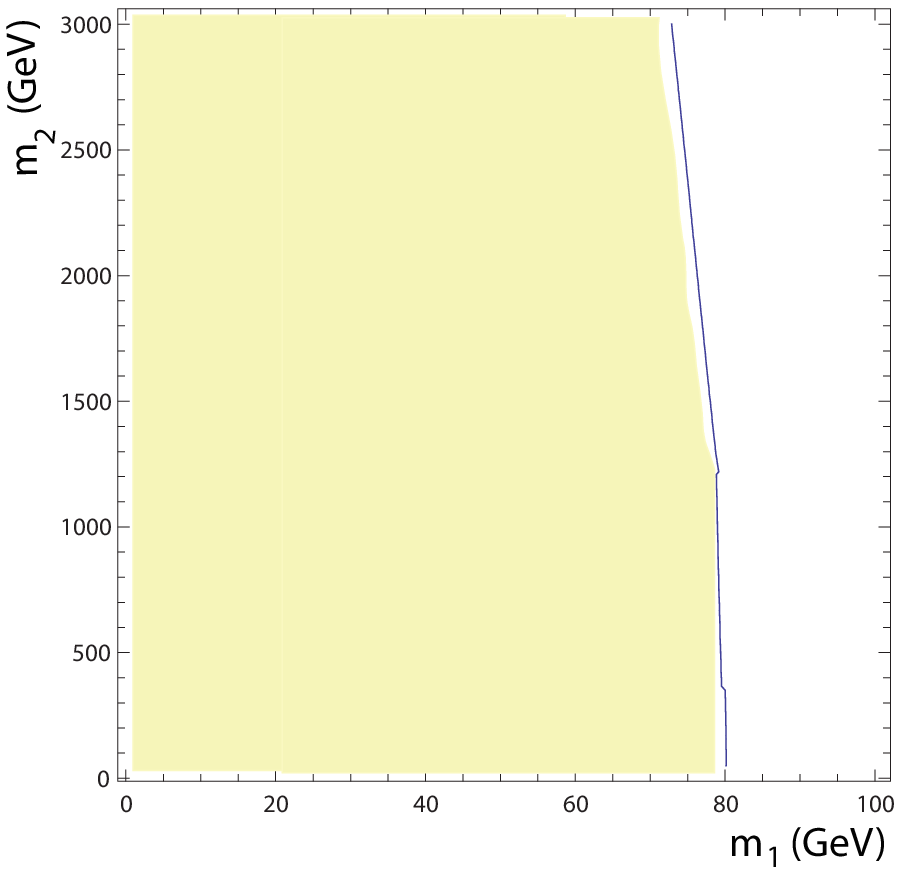} &
\includegraphics[width=1.5in]{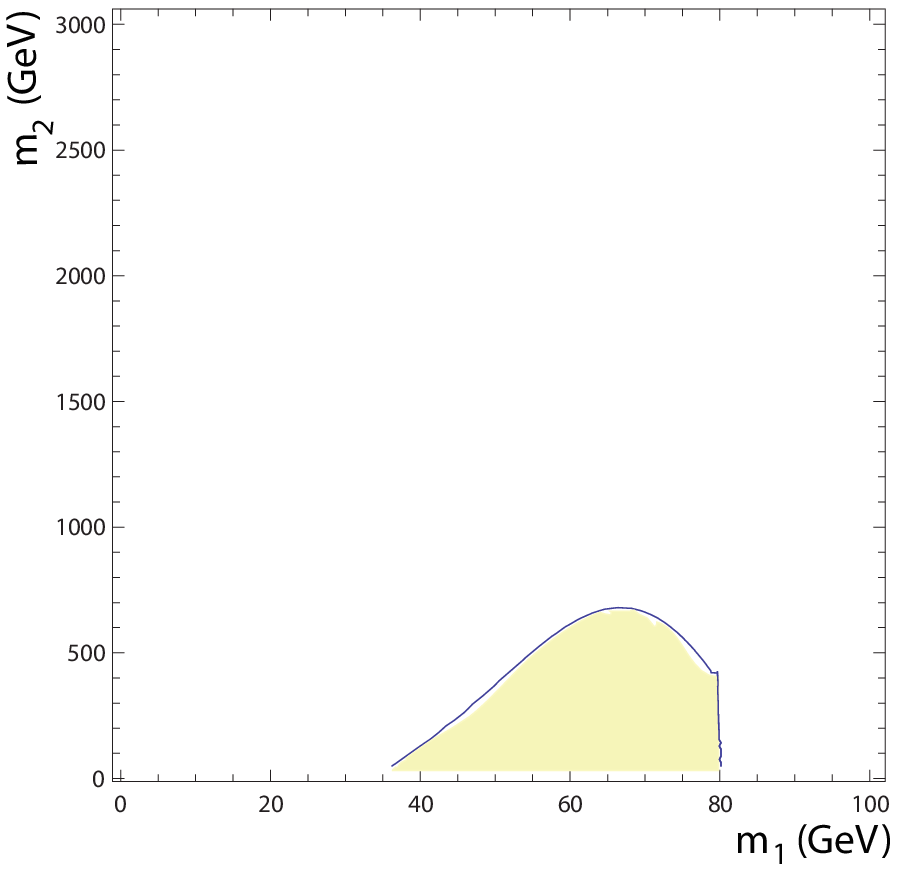}
\end{array}$
\end{center}
\caption{Plots of exclusion at 95 percent confidence in the $N_1$
  $N_2$ mass plane for Higgs masses (clockwise from top left) of 131, 150,200, and 180 GeV by Tevatron's inclusive like-sign dilepton search. }
\end{figure}

 \section{Conclusions}

 We have shown that Higgs production and decay plays a significant
 role in probing the existence of fourth generation leptons.  The
 existence of fourth generation quarks would yield a substantial
 increase in Higgs production from gluons fusion. In addition, the
 Higgs has large branching fractions into fourth generation neutrinos
 for Higgs masses up to about 500 GeV.  If the fourth generation
 neutrinos have both Dirac and Majorana states, we find that there are
 many Higgs decays into the channel $h\rightarrow N_1 N_1\rightarrow
 WW\ell\ell$ which yield many like-sign dilepton events.

We have calculated
 exclusion curves using CDF and ATLAS limit. If the heavy neutrinos
 decay exclusively to muons, the ATLAS inclusive like-sign muon analysis can exclude lightest neutrino masses between 100 and 200 GeV unless one is in the deep Majorana regime for fourth generation neutrinos.  In addition CDF's inclusive like-sign dilepton analysis excludes the lightest allowed fourth generation neutrinos in the case that the neutrino decay is flavor democratic.

 There are many prospects for fourth work along these lines.  The LHC
 experiments will soon publish results with nearly 1 fb$^{-1}$ of
 data.  Further investigation into like sign dileptons may rule out
 even more fourth generation parameter space.  However, if fourth
 generation neutrinos exist, there may be a surprising new channel in
 which to find the Higgs.   In this case, if the Higgs mass is below
 500 GeV,  a few percent of the Higgs branching fraction will be to
 fourth generation neutrinos, and hence to like-sign dileptons with
 many jets.  It may be possible to perform a search for the Higgs in
 this distinctive and low background region.  A similar search, with
 hard like-sign dileptons and jets was proposed to look for a fourth
 generation charged lepton and neutrino pair
 (\cite{Carpenter:2010bs}), one would expect in a fourth generation
 scenario with  Higgs production cross section around 10 picobarns,
 prospects would be quite good for Higgs discovery in this channel.

There is also a possibility for fourth generation scenarios which evade all current bounds and make for interesting Higgs physics; this is the case where the dominant branching fraction of fourth generation neutrinos is into taus.  Here the bound on fourth generation neutrino mass is quite light, only 61.2 GeV.  Here, for Higgs masses under 160 GeV it is still possible that the dominant Higgs decay is to fourth generation neutrinos; this is a new hidden Higgs scenario.  The dominant Higgs decay channel would then be $h\rightarrow N_1 N_1 \rightarrow WW\tau\tau$, beating the standard $h\rightarrow b\overline{b}$ signal by many orders of magnitude; this presents an interesting but difficult scenario for Higgs discovery.

{\bf Acknowledgments}

This work was supported in part by DOE grant number DE-FG03-92ER40689.

\end{document}